\documentclass[10pt]{iopart}
\usepackage[utf8]{inputenc}
\usepackage{hyperref}
\usepackage{graphicx}
\usepackage{siunitx}
\usepackage{xcolor}
\usepackage{xspace}
\usepackage{iopams}

\newcommand{\state}[3]			{{}^{#1}\mathrm{#2}_{#3}}						
\newcommand{\species}[3][]		{{}^{#1}\mathrm{#2}^{#3}}						
\newcommand{\transition}[2]		{#1 \mbox{--} #2}								

\newcommand{\ratio}				{\rho}											
\newcommand{\ratioYbSr}			{\rho_{\species{Yb}{+}/\species{Sr}{}}}			
\newcommand{\nuSr}				{\nu_{\species{Sr}{}}}							
\newcommand{\nuYb}				{\nu_{\species{Yb}{+}}}							
\newcommand{\adev}				{\sigma_y}										
\newcommand{\tavg}				{\tau}											
\newcommand{\redchisqr}         {\chi^2_\mathrm{red}}                           
\newcommand{\udark}             {u_\mathrm{d}}                                  
\newcommand{\wt}[1]             {w_{#1}}                                        
\newcommand{\ua}[1]             {u_{\mathrm{a}, #1}}                            
\newcommand{\ub}[1]             {u_{\mathrm{b}, #1}}                            

\newcommand{\textion}[2][+]		{#2\textsuperscript{#1}\xspace}					
\newcommand{\textisotope}[2][]	{\textsuperscript{#1}#2\xspace}					

\newcommand{\Sr}[1][]			{\textisotope[#1]{Sr}}							
\newcommand{\Yb}[1][]			{\textisotope[#1]{\textion{Yb}}}				
\newcommand{\FC}                {FC\xspace}                                     
\newcommand{\LO}                {LO\xspace}                                     
\newcommand{\TL}                {TL\xspace}                                     
\newcommand{\YbOne}				{Yb-1\xspace}									
\newcommand{\SiTwo}             {Si-2\xspace}                                   
\newcommand{\SrFC}				{FC(Sr)\xspace}		                		    
\newcommand{\YbLO}              {LO(Yb)\xspace}                                 
\newcommand{\SrLO}              {LO(Sr)\xspace}                                 
\newcommand{\YbTL}              {TL(Yb)\xspace}                                 
\newcommand{\SrTL}              {TL(Sr)\xspace}                                 

\begin{document}
\title{
	Optical frequency ratio of a \textsuperscript{171}Yb\textsuperscript{+} single-ion clock and a \textsuperscript{87}Sr lattice clock
}

\author{
    S~D\"orscher, N~Huntemann, R~Schwarz, R~Lange, E~Benkler, B~Lipphardt, U~Sterr, E~Peik, and C~Lisdat
}
\address{
	Physikalisch-Technische Bundesanstalt, Bundesallee 100, 38116 Braunschweig, Germany
}
\ead{
	soeren.doerscher@ptb.de
}

\vspace{10pt}
\begin{indented}
	\item[]4 September 2020
\end{indented}

\begin{abstract}
	We report direct measurements of the frequency ratio of the \SI{642}{\tera\hertz} $\transition{\state{2}{S}{1/2} (F=0)}{\state{2}{F}{7/2} (F=3)}$ electric octupole transition in \Yb[171] and the \SI{429}{\tera\hertz} $\transition{\state{1}{S}{0}}{\state{3}{P}{0}}$ transition in \Sr[87].
	A series of 107 measurements has been performed at the Physikalisch-Technische Bundesanstalt between December 2012 and October 2019.
	Long-term variations of the ratio are larger than expected from the individual measurement uncertainties of few \num{e-17}.
	The cause of these variations remains unknown.
	Even taking these into account, we find a fractional uncertainty of the frequency ratio of \num{2.5e-17}, which improves upon previous knowledge by one order of magnitude.
	The average frequency ratio is $\nuYb / \nuSr = \num{1.495 991 618 544 900 537(38)}$.
	This represents one of the most accurate measurements between two different atomic species to date.
\end{abstract}
	
%
\vspace{2pc}
\submitto{\MET}
%
%
\ioptwocol
%


\section{Introduction}
\label{sec:intro}
The International System of Units (SI) has been revised fundamentally in 2019, redefining several of its base units \cite{bip19}.
While the definition of the SI base unit of time, the `second', has remained essentially unchanged, there has been tremendous progress in time and frequency metrology over the past decades \cite{biz19}. 
In particular, optical atomic clocks have surpassed primary realizations of the second in terms of frequency instability \cite{sch17, oel19, sch20d} and systematic uncertainty \cite{mcg18, san19, bre19, bot19} by two orders of magnitude or more.
Thus, the International Committee for Weights and Measures recommends standard frequency values of several atomic transitions \cite{rie18} as Secondary Representations of the Second (SRS) \cite{cct04}.
They presently include nine transitions with fractional uncertainties between \num{4E-16} and \num{2E-15}; eight of these are optical transitions  \cite{rie18}.
Recently, measurements of transition frequencies with uncertainties near and even below \num{2E-16} have been reported for transitions in optical clocks using \textisotope[171]{Yb} \cite{mcg19} and \Sr[87] \cite{sch20d, nem20}.
These absolute frequency measurements are limited by the uncertainty of the \textisotope[133]{Cs} primary frequency standards (PFS) serving as the reference. 
Since the ratio $\ratio$ of any two unperturbed atomic clock transition frequencies is a universal, dimensionless constant, the direct determination of an optical frequency ratio via a frequency comb \cite{ste02a} permits significantly smaller uncertainty. 

The Working Group of Strategic Planning of the Consultative Committee for Time and Frequency has presented a roadmap towards a redefinition of the second based on an optical transition that outlines the guiding requirements on optical clocks \cite{rie18}. 
Measurements of optical clock frequency ratios with uncertainties below \num{5E-18} are among the milestones of this roadmap. 
Furthermore, direct intercomparisons are essential for applications in fundamental research, such as searches for variations of fundamental constants \cite{god14, hun14} or indications of dark matter \cite{rob20}, and in applied sciences, e.g., in geosciences for measuring geopotential differences \cite{gro18a, tak20}.

Various means to compare the optical clocks of different laboratories have been investigated.
Phase-stabilized optical fibre links and satellite-based links permit remote frequency ratio measurements over continental \cite{lis16} and intercontinental distances \cite{fuj18}, respectively. 
Transportable optical clocks \cite{kol17, cao17, tak20} enable comparisons without direct interconnection between laboratories. 
Another approach is based on measuring the frequency ratio of a heterogeneous pair of clocks ($\ratio \neq 1$) locally.
The ratios determined at different laboratories can be reported and subsequently compared to evaluate the clock performances. 
Moreover, closures of the pairwise frequency ratios between three or more types of clocks can be used to test reproducibility, even if only a single laboratory operates a specific pair.
Evaluations of closures at the level of \num{e-16} fractional uncertainty have been reported recently \cite{mcg19, nem20, ohm20}.
Due to the correlations between optical frequency ratio measurements, new methods \cite{mar15, rob16a} have emerged for their combined evaluation, e.g., in the context of SRS \cite{rie18}.

Comparisons of optical clocks using different species with uncertainty below \num{1E-16} have been reported by several groups \cite{ros08, tak15, yam15, nem16, ori18, ohm20, bel20a}.
Here, we report on direct comparisons of a \Yb[171] single-ion clock realizing the $\transition{\state{2}{S}{1/2} (F=0)}{\state{2}{F}{7/2} (F=3)}$ electric octupole (E3) transition frequency $\nuYb \approx \SI{642}{\tera\hertz}$ and a \Sr[87] optical lattice clock interrogating the $\transition{(5s^{2})\,\state{1}{S}{0}}{(5s5p)\,\state{3}{P}{0}}$ transition at $\nuSr \approx \SI{429}{\tera\hertz}$.
A series of measurements with uncertainties of few \num{e-17} has been conducted at the  Physikalisch-Technische Bundesanstalt (PTB), spanning a period of seven years.
These are the first measurements reported for this specific pair of clock species to date.
We investigate the reproducibility of our measurements and derive an average value of the frequency ratio.

\section{Experimental setup}
\label{sec:setup}
We compare the \Yb single-ion clock \YbOne and the \Sr optical lattice clock at PTB and measure their frequency ratio $\ratioYbSr = \nuYb / \nuSr$ directly by means of an optical frequency comb.
The clocks are located in different buildings at a distance of about \SI{200}{\meter} on the PTB campus.
\Fref{fig:setup} illustrates the experimental setup; a detailed description is given below.
\begin{figure*}
    \centering
    \includegraphics{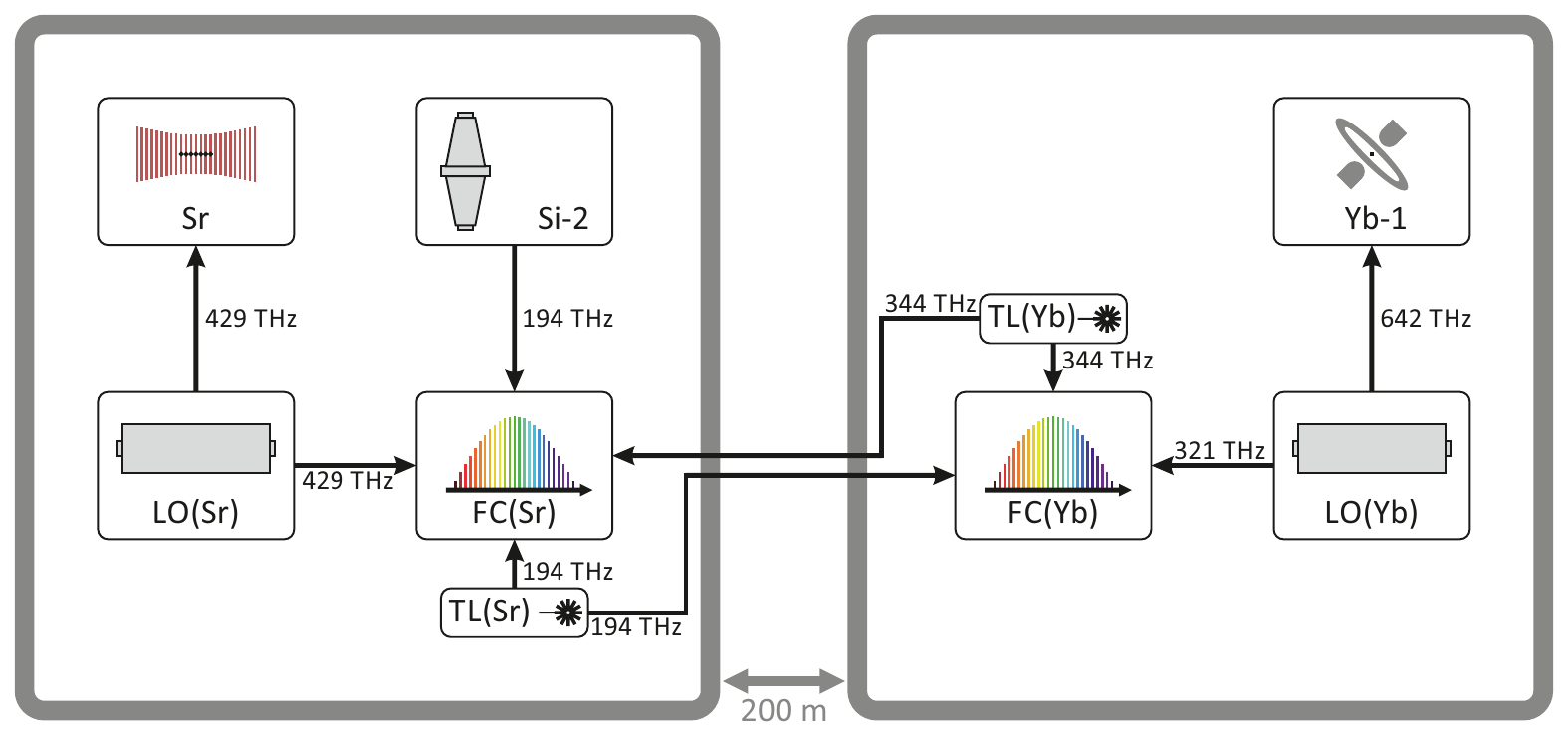}
    \caption{
        Experimental setup for comparing the \YbOne single-ion clock and the \Sr lattice clock at PTB.
        The clocks are located in different buildings, about \SI{200}{\meter} apart.
        The measurement topology is illustrated schematically at the level of key components (\FC: frequency comb, \LO: local oscillator, \TL: transfer laser) in each building, including the \SiTwo ultrastable laser system.
        Black lines indicate phase-stabilized fibre links.
        A detailed description is given in \sref{sec:setup} of the main text.
    }
    \label{fig:setup}
\end{figure*}

\subsection{Frequency combs and links}
\label{sec:setup:combs}
In each building, an optical frequency comb (\FC) is linked by an optical fibre to the local oscillator (\LO) of the respective clock. 
To establish a phase-coherent link between the two frequency combs, light from transfer lasers, \YbTL and \SrTL, operating near \SI{344}{\tera\hertz} and \SI{194}{\tera\hertz}, is sent to both combs. 
All fibre links shown in \fref{fig:setup} are actively phase-stabilized. 
Using the transfer oscillator scheme \cite{ste02a}, we determine the frequency ratio $\ratioYbSr$ from radio-frequency (rf) beat signals. 
Here, we typically employ the \SrTL transfer laser and use \YbTL to validate link and comb performance during measurements.

Dead-time-free counters record the frequencies of all rf signals required to determine $\ratioYbSr$.
They are operated in phase-averaging mode ($\Lambda$-mode) with a gate time of \SI{1}{\milli\second}, a report time of \SI{1}{\second}, and synchronized to a common pulse-per-second (PPS) signal.
Synchronous acquisition avoids frequency offsets caused by drifts of the acquired signals and rejects noise from the transfer lasers and combs.
The PPS signal and a \SI{100}{\mega\hertz} reference are transmitted from the building housing the \Yb clock to the one housing the \Sr clock, using high-quality rf cables (see also \cite{sch20d}).

We monitor the phase stabilization of the fibre paths to and between the combs by counting the respective in-loop beat frequencies, as phase fluctuations on the paths enter directly into the calculation of $\ratioYbSr$.
Some of the rf signals are band-pass-filtered by electronic tracking with phase-locked voltage-controlled oscillators before being counted.
In these cases, we track and count two replicas with different settings of the phase-locked loops.
Differences of the two frequencies indicate cycle slips of a tracking oscillator.
Invalid data are filtered out in post-processing.

\subsection{\Yb single-ion clock}
\label{sec:setup:yb}
The ytterbium ion clock \YbOne at PTB has been described in detail in previous publications \cite{hun12, hun14, hun16, san19}.
It realizes the frequency of the $\transition{\state{2}{S}{1/2} (F=0)}{\state{2}{F}{7/2} (F=3)}$ E3 transition  using a single \Yb[171] ion confined in a radio-frequency Paul trap. 
The clock has recently been evaluated with a fractional systematic uncertainty of \num{2.7E-18} \cite{san19}.
During the measurements reported here, the clock has been operated either with this systematic uncertainty or the slightly larger uncertainty reported in \cite{hun16}.

\begin{figure}
    \centering
    \includegraphics{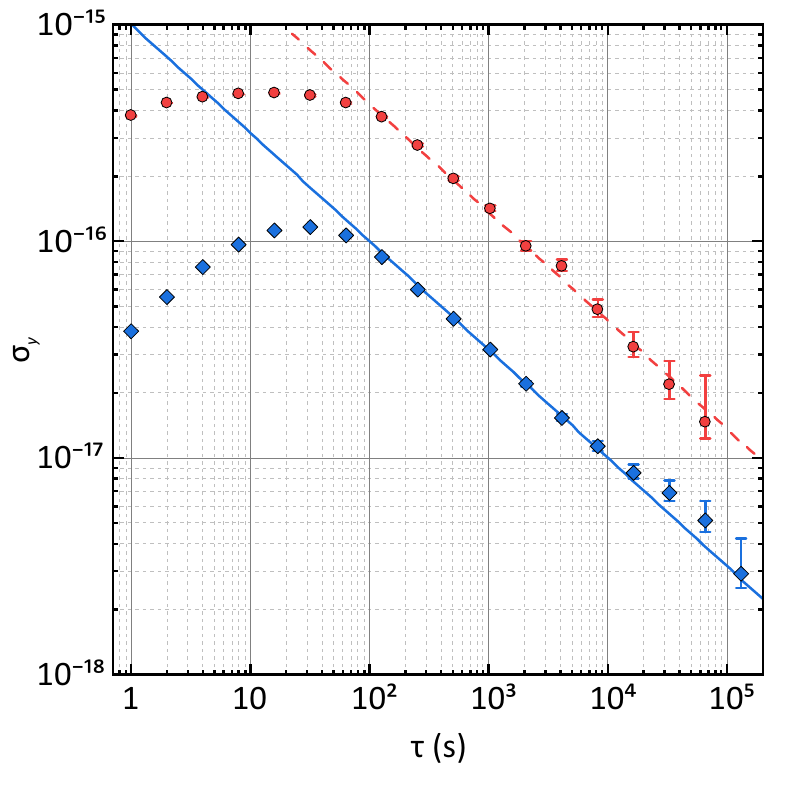}
    \caption{
        Overlapping Allan deviation $\adev(\tavg)$ of a typical frequency ratio measurement with (blue diamonds) and without (red circles) stability transfer from the \SiTwo ultrastable laser to the \LO of the \Yb and \Sr clocks.
        Fractional frequency instabilities of $\num{1.0e-15} (\tavg / \si{\second})^{-1/2}$ (solid blue line) and $\num{4.3e-15} (\tavg / \si{\second})^{-1/2}$ (red dashed line) are observed.
        The underlying data sets are those from Modified Julian Dates 57906 through 57928 and 57184 through 57197, respectively.
    }
    \label{fig:oadev}
\end{figure}

Since early 2017, the noise properties of the probe laser are significantly improved by stabilizing it to a laser system referenced to the length of a cryogenic silicon cavity (\SiTwo) \cite{mat17a}. 
As a result, the fractional instability of the frequency ratio as expressed by the Allan deviation (ADEV) $\adev(\tavg)$, where $\tavg$ is the averaging time, reduces from $\adev(\tavg) \approx \num{4.3E-15} (\tavg / \si{\second})^{-1/2}$ to $\adev(\tavg) \approx \num{1.0E-15} (\tavg / \si{\second})^{-1/2}$ (see \fref{fig:oadev}).

\subsection{\Sr lattice clock}
\label{sec:setup:sr}
The laboratory strontium lattice clock at PTB has been described in previous publications \cite{fal14, alm15, gre16, lis16, sch20d}.
We interrogate the $\transition{(5s^{2})\,\state{1}{S}{0}}{(5s5p)\,\state{3}{P}{0}}$ transition of a few hundred laser-cooled \Sr[87] atoms that are confined in a nearly horizontal optical lattice.

All measurements reported in this article have been performed using one apparatus, which underwent the following modifications:
We replaced the plate capacitor, which had been used to measure the static differential polarizability of the \Sr clock transition \cite{mid12a}, by a copper structure, into which the atomic sample can be transported for interrogation with well-controlled BBR (`cold environment') \cite{mid11}, in late 2014.
However, it was not used for any of the measurements reported here.
Its temperature was monitored and included in the evaluation of the BBR field.
We removed it in early 2018 (see \sref{sec:data_analysis:checks:sr}).
Several laser systems have been replaced over time \cite{hae15a, sch19b}.
We have been using the same titanium:sapphire lattice laser system since 2014.
Its output is spectrally filtered by a volume Bragg grating, which has a full width at half maximum of about \SI{60}{\giga\hertz}, as a precaution to avoid light shifts due to amplified spontaneous emission (ASE).

\begin{table}
    \caption{
        \label{tab:unc:sr}
        Typical uncertainty budget of the \Sr lattice clock in low-uncertainty mode, as reported recently in \cite{sch20d}.
        All values are given in units of $\nuSr$.
    }
    \begin{indented}
        \item[]
        \begin{tabular}{lSS[table-format=<1.2]}
            \br
            \textrm{Effect}&
                \multicolumn{1}{c}{\textrm{Frequency shift}}&
                \multicolumn{1}{c}{\textrm{Uncertainty}}\\
            &
                \multicolumn{1}{c}{(\num{e-18})}&
                \multicolumn{1}{c}{(\num{e-18})}\\	
            \mr
            Optical lattice             &    -3.6   &  3.1  \\
            Blackbody radiation         & -4906.4   &  13.7 \\
            Blackbody radiation (oven)  &    -3.0   &  1.2  \\
            Quadratic Zeeman            &  -134.1   &  1.0  \\
            Cold collisions             &    -0.6   &  0.9  \\
            Servo error                 &     0.0   &  0.7  \\
            Tunnelling                  &     0.0   &  4.8  \\
            DC Stark                    &    -2.0   &  0.7  \\
            Background gas collisions   &    -1.9   &  1.9  \\
            Other                       &     0.0   & <0.1  \\
            \mr
            Total                       & -5052     & 15    \\
            \br
        \end{tabular}
    \end{indented}
\end{table}

The fractional systematic uncertainty of the strontium clock has been improved gradually from \num{3E-17} \cite{fal14} to about \num{1.5E-17} \cite{sch20d} (see \tref{tab:unc:sr}).
As improved interrogation laser systems \cite{hae15a, mat17a} have become available, we frequently operate the clock with an additional dead time of one to two seconds between subsequent interrogation cycles (`low-uncertainty mode').
This reduces the thermal load on the physics package and results in decreased blackbody radiation (BBR) uncertainty \cite{sch20d}, while preserving an acceptable frequency instability (see below).
In the current apparatus, this uncertainty is limited both by the temperature inhomogeneity of the vacuum system itself and by the temperature gradient along the main magnetic field coils, which are water-cooled and located inside the vacuum system.
We have also improved our characterization of the frequency shift due to BBR from the strontium oven by a factor of eight, using a mechanical shutter for a direct measurement of the differential effect \cite{sch20d}.
The improved characterization has been applied retroactively to earlier measurements.
Moreover, we have reduced the uncertainty of optical lattice light shifts by a factor of three \cite{sch20d}, using a procedure similar to the `operational magic wavelength' \cite{kat15} to better account for correlations of light shift contributions.
We use a light-shift model that accounts for a possible intensity imbalance between the lattice beams due to reflection losses.
It is equivalent to the model presented in \cite{nem19}.

The strontium clock's frequency instability has been improved from about $\adev(\tavg) \approx \num{6E-16} (\tavg / \si{\second})^{-1/2}$ to $\adev(\tavg) \approx \num{2E-16} (\tavg / \si{\second})^{-1/2}$ \cite{alm15} by an improved local oscillator \cite{hae15a}.
It has been reduced further, to $\adev(\tavg) \approx \num{5E-17} (\tavg / \si{\second})^{-1/2}$ \cite{sch20d}, by transferring the stability of \SiTwo \cite{mat17a} to \SrLO in 2016.
These values refer to operating the clock without additional dead time (`high-stability mode').
In low-uncertainty mode, frequency instability is higher, but still nearly an order of magnitude below that of the \Yb clock.

For the measurements reported here, the \Sr clock has usually been operated in low-uncertainty mode, with fractional systematic uncertainties typically between \num{1.5E-17} and \num{2E-17}.
We also include some data from high-stability operation, e.g., the monitoring of lattice light shifts.
Typically, the uncertainty of these measurements has been only slightly larger, mainly between \num{1.8e-17} and \num{3.0e-17}.

\subsection{Gravitational redshift correction}
\label{sec:setup:geo}
The fractional gravitational redshift between the \Sr lattice clock and \YbOne has been determined as \num{69.5(5)e-18} \cite{den17}.

\section{Data analysis}
\label{sec:data_analysis}

\begin{figure*}
    \centering
    \includegraphics{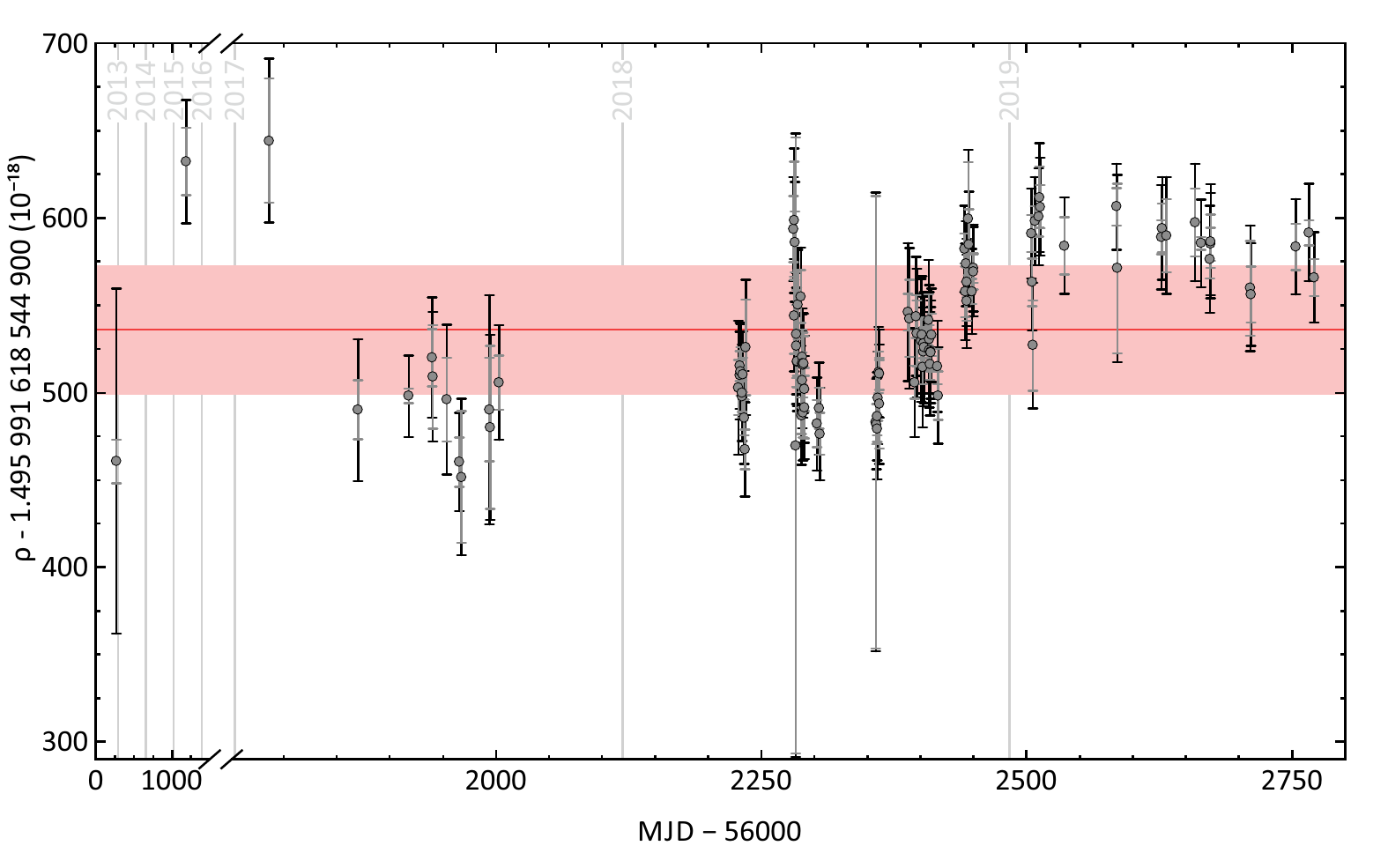}
    \caption{
    Values of the frequency ratio $\ratioYbSr$ (grey-filled circles) measured in direct comparisons of the \YbOne single-ion clock, realizing the E3 transition, and the laboratory \Sr lattice clock at PTB, by Modified Julian Date (MJD) of the measurement.
    Grey and black error bars indicate the statistical and total measurement uncertainties, respectively.
    The solid red line and red-shaded area illustrate the average frequency ratio and standard uncertainty derived in \sref{sec:data_analysis:avgratio}.
    Vertical, light grey lines indicate the MJD corresponding to 1 January from 2013 through 2019.
    }
    \label{fig:measurements}
\end{figure*}

We have performed a series of frequency comparisons between the \Yb clock and the \Sr clock directly in the optical domain, as discussed in \sref{sec:setup}.
This series consists of a total of 107 measurements recorded between December 2012 and October 2019.
\Fref{fig:measurements} shows the measured values of the frequency ratio $\ratioYbSr$ by date.
The improved frequency stability of \YbOne since 2017 has enabled an increased rate of measurements, because it reduces the typical measurement duration required for a statistical uncertainty of about \num{1e-17} to a few hours.
This has been particularly useful for the investigations discussed in \sref{sec:data_analysis:checks}.

The measurements shown in \fref{fig:measurements} are typically separated by interruptions of considerable duration or by deliberate changes of the operating conditions of either clock;
brief gaps within a data set are tolerated.
In total, we have recorded more than \SI{1000}{\hour} of data.

The fractional uncertainty of an individual measurement is typically about \num{2.5e-17}, with comparable contributions from the systematics of the \Sr clock and statistics.
Notable exceptions include the first measurement from 2012, which had a higher systematic uncertainty than later measurements, and several brief measurements, which are dominated by the statistical uncertainty.

As shown in \fref{fig:measurements}, the scatter of our measurements is typically no more than a few parts in \num{e17} on short time scales, but may be up to about one part in \num{e16} over periods of half a year or longer.
For data sets up to a few days, we compute the ADEV of the frequency ratio (see \fref{fig:oadev}) and observe no significant deviation from white frequency noise behaviour beyond the initial servo attack time.
For the full data set, we find a reduced chi-square $\redchisqr = 1.9$ with respect to the total uncertainties, however.
This shows that long-term scatter of the ratio seen in \fref{fig:measurements} is larger than expected from the uncertainties of the individual measurements.

\subsection{Investigation of potential unaccounted frequency shifts}
\label{sec:data_analysis:checks}
The unexpectedly large long-term variations of the ratio may be caused by an incorrectly characterized or unknown systematic frequency shift.
We have thus been investigating the performance of both clocks as well as that of the links and combs to exclude or confirm potential sources of error, since the first significant variation of the ratio was observed in early 2017.

We have carefully checked the experimental setup and our evaluation of the systematic frequency shifts.
Where possible, we have not only reevaluated systematic effects by interleaving clock stabilizations at different settings in a single clock, but also by monitoring the frequency ratio $\ratioYbSr$.
Moreover, we have deliberately changed parts of the experimental setup to verify that they have no influence on the measured frequency ratio.

\subsubsection*{\Yb clock}
\label{sec:data_analysis:checks:yb}
The following effects have been investigated for the single-ion clock:
\begin{itemize}
    \item
    \emph{Excess micromotion} in the \Yb clock can be caused by uncompensated DC electric fields at the position of the ion and results in second-order Doppler and Stark shifts.
    It is monitored by analyzing the fluorescence light using the photon correlation technique \cite{kel16,ber98} during the cooling period of the clock cycle.
    Typically, we minimize it as discussed in \cite{hun16} once per day.
    Because of the large mass of the ytterbium ion, large intentionally applied deviations from the optimal configuration of compensation fields only lead to expected fractional frequency shifts on the order of \num{e-17}, marginally resolvable in the frequency ratio.
    
    \item
    \emph{Light-shift cancellation} is important for the realization of an optical clock based on the E3 transition.
    For all measurements reported here, we use a hyper-Ramsey interrogation sequence \cite{yud10} for which we control the step frequency that compensates the light shift by Rabi spectroscopy in interleaved interrogation cycles \cite{hun16}.
    To investigate for potential limitations of the light-shift cancellation, the duration of Ramsey pulses has been reduced by a factor of two and consequently the light shift during the pulses by a factor of four.
    Furthermore, the free atomic state evolution between Ramsey pulses has been changed from \SI{60}{\milli\second} up to \SI{450}{\milli\second}.
    
    \item
    The \emph{ion trap drive rf amplitude} as well as the \emph{orientation and magnitude of the externally applied magnetic field} have mainly been kept constant, but have been varied deliberately for some measurements.
    
    \item
    The \emph{local oscillator \YbLO} has been checked for as a possible source of frequency shifts by disabling the prestabilization to \SiTwo for some measurements.
\end{itemize}

\subsubsection*{\Sr clock}
\label{sec:data_analysis:checks:sr}
The following effects and systems have been investigated for the lattice clock:
\begin{itemize}
    \item
    The \emph{DC Stark shift} due to residual electric fields has been checked regularly by applying an external electric field as described in \cite{gre16}.
    Moreover, the magnetic field coils effectively shield the atomic sample from its environment, especially, since the cold environment has been removed (see \sref{sec:setup:sr}).
    The electric potentials of the coils themselves were measured directly.
    
    \item
    The \emph{lattice light shift} has been monitored for potential inconsistency with our model, e.g., due to a residual ASE background.
    We have varied the operating conditions of the lattice laser (see \sref{sec:setup:sr}), such as the pump laser power.
    A change in the lattice laser frequency due to the improved treatment of lattice light shifts (see \sref{sec:setup:sr}) serves as an additional test for such inconsistencies.
    
    \item
    \emph{Stray light} has been checked as a potential source of additional light shifts, by selectively blocking light sources or by installing additional mechanical shutters in laser systems.
    
    \item
    The \emph{quadratic Zeeman shift} has been checked by varying the external magnetic field during spectroscopy by a factor of five.
    
    \item
    \emph{BBR from the \Sr oven} has been characterized by measuring the frequency shift directly (see \sref{sec:setup:sr} and \cite{sch20d}).
    
    \item
    \emph{Collisions with the background gas}, which may cause frequency shifts, have varied over the course of the measurement series, due to changes in residual pressure.
    Measured trap lifetimes range from below \SI{1}{\second} through more than \SI{5}{\second}.
    Additionally, we have installed a mass spectrometer to monitor the composition of the background gas.
    
    \item
    \emph{Line pulling} has been checked for by varying the duration of the spectroscopy pulse by about a factor of two.
    
    \item
    The \emph{local oscillator \SrLO} has been checked for as a possible source of frequency shifts.
    We have disabled the prestabilization to \SiTwo for some measurements.
    Likewise, the phase stabilization from \SrLO to the atomic sample has been disabled for other measurements, to check for indications for uncompensated phase changes.
    The latter measurements are not included in the data set analyzed here.
    
    \item
    We have also checked the \emph{control system} of the clock apparatus by introducing a variable delay between preparation and interrogation, by toggling the synchronization of the experimental sequence to the mains phase, and by changing the order of the four interrogations from which the error signal is derived \cite{fal14}.
\end{itemize}

\subsubsection*{Frequency combs}
\label{sec:data_analysis:checks:combs}
The investigations regarding the frequency combs include the following tests and modifications:
\begin{itemize}
    \item
    The \emph{fractional frequency offset and instability} of the comb-to-comb connection between the two buildings including contributions from the combs themselves are assessed regularly.
    The difference between the fractional frequency ratios of the transfer lasers \YbTL and \SrTL measured with the two combs is  smaller than \num{2e-18} for all data sets longer than \SI{5e3}{\second}, with typical fractional instabilities of \num{3e-17} at $\tavg = \SI{1}{\second}$, \num{3e-18} at $\SI{e3}{\second} \le \tavg \le \SI{e4}{\second}$, and \num{2e-19} at $\tavg = \SI{e6}{\second}$.
    All these quantities are well below the uncertainty and instability of the fractional frequency ratio of the two clocks at the respective averaging times.

    \item
    \emph{Evaluation using a single frequency comb} has been performed to exclude any considerable contributions of the links and frequency combs to the instability and uncertainty of the frequency ratio measurement.
    To this end, the light field of \YbLO has been sent to \SrFC via a fibre link for some of the measurements in 2018.
    The frequency ratio $\ratioYbSr$ was determined using only \SrFC, in addition to the method outlined in \sref{sec:setup:combs}.
    Both results agreed on the level of the uncertainty and instability as evaluated above using \YbTL and \SrTL.
\end{itemize}

We have not found any inconsistencies with our evaluation of the systematic shifts or any significant correlation between recorded changes made to the experimental setup and the frequency ratio in any of these measurements.

\subsection{Average frequency ratio}
\label{sec:data_analysis:avgratio}
As discussed in \sref{sec:data_analysis:checks}, the cause of the observed long-term variations of the frequency ratio $\ratioYbSr$ remains unknown.
Larger scatter than expected from the measurement uncertainty has also been observed in other comparisons involving optical clocks \cite{bai08, bel20a}.
Similarly, we treat the excess scatter as the result of a varying, unknown systematic frequency shift with a constant, `dark' uncertainty $\udark$.
This approach implicitly assumes that the shift scatters around zero rather than an unknown offset.

Our measurements strongly suggest that this shift is predominantly varying slowly in time (see \fref{fig:measurements}) and thus correlated between measurements in temporal proximity.
We determine $\udark / \ratioYbSr = \num{18e-18}$ by requiring that $\redchisqr = 1$ for our data set if $\udark$ is included as an additional constant uncertainty in each measurement.
The average ratio of $\ratioYbSr$ is the arithmetic mean with weights $\wt{i}$ derived from the statistical uncertainty $\ua{i}$, the combined systematic uncertainty $\ub{i}$ of the clocks, and the common dark uncertainty $\udark$ for each measurement, as $\wt{i}^{-1} = \ua{i}^2 + \ub{i}^2 + \udark^2$.
The uncertainty of the average frequency ratio is the root-sum-square value of the contributions from statistics, systematics, and the unknown effect. 
The former is below \num{2e-18} for the full data set.
Systematic shifts and the unknown effect are each treated as correlated.
Hence, their combined uncertainty is not reduced by averaging;
it is determined through the average value of the variances $\ub{i}^2 + \udark^2$ using the weights $\wt{i}$.
We find an average frequency ratio $\ratioYbSr = \num{1.495 991 618 544 900 537(38)}$, as shown in \fref{fig:measurements}.

\begin{figure}
    \centering
    \includegraphics{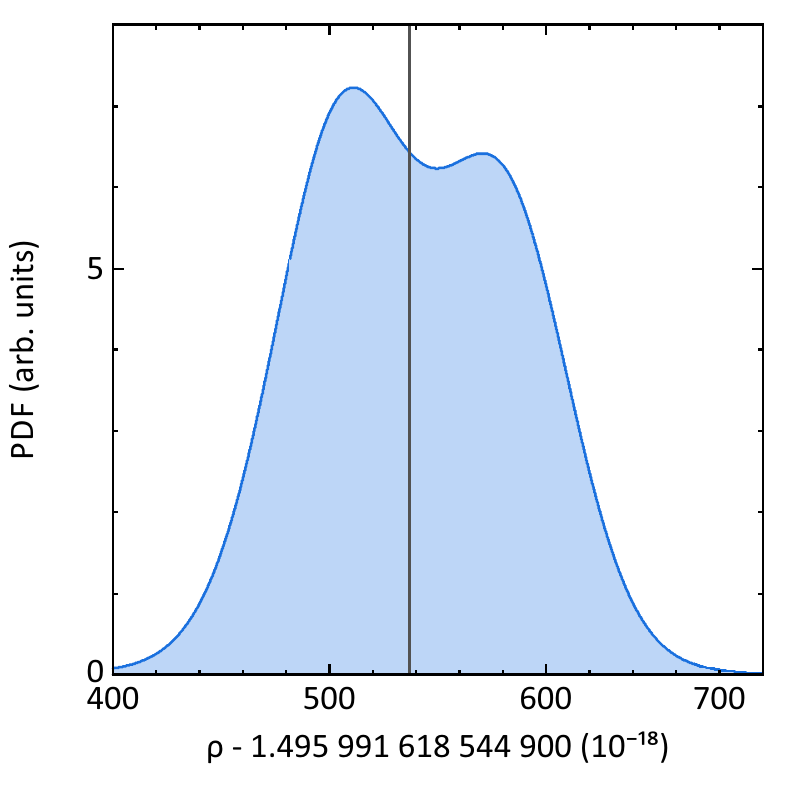}
    \caption{
        Probability density function resulting from aggregating the measurements shown in \fref{fig:measurements} (blue, shaded curve) as discussed in the main text.
        The mean value of the distribution is shown as a grey line.
    }
    \label{fig:linpool}
\end{figure}

We have compared this result to those of other methods to confirm that the value found for $\ratioYbSr$ is robust and its uncertainty reproducible.
Scaling the individual measurements' uncertainties by $\sqrt{\redchisqr}$, similar to \cite{bel20a}, leads to an average value that deviates from the above result by \num{-2e-19} and an uncertainty of \num{3.5e-17}.
\Fref{fig:linpool} shows the result of aggregating the individual measurements, similar to the linear pooling method for finding a consensus of independent measurements \cite{koe17, koe17a}.
The mixed probability density function (PDF) is the sum of normal distributions that are centered around the measured value and that have a second moment $\sqrt{\ua{i}^2 + \ub{i}^2}$ and weight $1 / \left(\ua{i}^2 + \ub{i}^2 \right)$ for each measurement.
The resulting mean value of $\ratioYbSr$ deviates from the result above by \num{2e-18}.
The uncertainty of \num{5.0e-17} resulting from the second moment of the mixed PDF is slightly higher than the uncertainty found above.
However, this method does not account for the reduction of the statistical uncertainty by averaging the individual measurements.

\section{Conclusion}
\label{sec:conc}
A series of comparisons between a \Yb[171] (E3) single-ion clock and a \Sr[87] optical lattice clock spanning a period of seven years has been performed at PTB.
The frequency ratio $\ratioYbSr$ of the undisturbed clock transitions has been measured with individual measurement uncertainties of few \num{e-17}.
These measurements demonstrate how the increasing stability and robustness of the clocks have enabled more frequent comparisons of the systems.

Our measurements have also revealed that the frequency ratio of the clocks, while well-reproducible within the clocks' combined uncertainty in the short term, exhibits an unexpectedly large scatter in the long term.
These variations are on the order of several \num{e-17}, which is below the resolution of absolute frequency measurements but exceeds the combined uncertainty of the clocks ($\redchisqr \lessapprox 2$).
In the present case, no uncontrolled frequency shifts could be identified despite an extensive search.
While this result is specific to the clocks used here, it highlights the importance of direct comparisons of optical clocks over extended periods of time or using multiple pairs of clocks.

Taking the long-term variation into account, we find a frequency ratio of $$\ratioYbSr = \num{1.495 991 618 544 900 537(38)}$$ between the clocks.
This result can be compared to future measurements of $\ratioYbSr$ at different laboratories, either directly or indirectly as part of a closure, to investigate the reproducibility in the combined system and to update the recommended frequency values of the SRS \cite{rie18}.
The fractional uncertainty of \num{2.5e-17} of our measurement improves upon knowledge from recent absolute frequency measurements \cite{sch20d, nem20, hun14} by an order of magnitude.
Furthermore, it is one of the lowest reported for comparisons of different clock species; a lower uncertainty has only been achieved in \cite{bel20a}.
To our knowledge, this is the only series of measurements directly comparing optical clocks of different type over seven years or longer.

\ack{
    We thank T~Legero, D~Matei, and S~H\"afner for operating the \SiTwo laser system, C~Grebing for operating one of the optical frequency combs during the first few measurements, and N~Lemke, S~Falke, A~Al-Masoudi, M~Abdel Hafiz, and C~Sanner for their contributions to operating the clocks during some measurements.
	We acknowledge support by the project 18SIB05 ROCIT, which has received funding from the EMPIR programme co-financed by the Participating States and from the European Union’s Horizon 2020 Research and Innovation Programme, and by the Deutsche Forschungsgemeinschaft (DFG, German Research Foundation) under Germany’s Excellence Strategy -- EXC-2123 QuantumFrontiers -- 390837967 and CRC~1227 DQ-\textit{mat} within project B02.
}

\section*{References}
\bibliographystyle{iopart-num}
\providecommand{\newblock}{}

\end{document}